\documentclass[12pt]{iopart}
\usepackage{iopams}
\usepackage{mathrsfs}
\usepackage{graphicx}
\usepackage{color}
\usepackage[english]{babel}
\usepackage{float}
\usepackage{upgreek}
\usepackage{dsfont}
\usepackage{sublabel}
\usepackage{cite}
\newcommand{\cL}{{\mathcal L}}

\newcommand{\cO}{{\mathcal O}}

\newcommand{\cQ}{{\mathcal Q}}

\newcommand{\ZZ}{{\mathbb Z}}

\newcommand{\sA}{{\mathsf A}}
\newcommand{\sB}{{\mathrm B}}
\newcommand{\sC}{{\mathrm C}}
\newcommand{\sD}{{\mathrm D}}

\newcommand{\sH}{{\mathsf H}}

\newcommand{\sJ}{{\mathsf J}}

\newcommand{\sR}{{\mathrm R}}

\newcommand{\pop}{{\mathsf p}}

\newcommand{\matel}[3]{{\left\langle \vphantom{#1 #2 #3} #1 \,\right\vert
\left.
 \hspace{-0.15em} \vphantom{#1 #2 #3} #2 \,\right\vert \left.
 \hspace{-0.15em} \vphantom{#1 #2 #3} #3\right\rangle}}

\newcommand{\dint}[1]{\mathrm{d} #1 ~}

\newcommand{\eqref}[1]{{(\ref{#1})}}
\begin{document}
 
\title{Quantum Angular Momentum Diffusion of Rigid Bodies}

\author{Birthe Papendell, Benjamin A. Stickler, and Klaus Hornberger}
\address{Faculty of Physics, University of Duisburg-Essen, Lotharstra\ss e 1, 47048 Duisburg, Germany}

\begin{abstract}
We show how to describe the diffusion of the quantized angular momentum vector of an arbitrarily shaped rigid rotor as induced by its collisional interaction with an environment. We present the general form of the Lindblad-type master equation and relate it to the orientational decoherence of an asymmetric nanoparticle in the limit of small anisotropies. The corresponding diffusion coefficients are derived for gas particles scattering off large molecules and for ambient photons scattering off dielectric particles, using the elastic scattering amplitudes.
\end{abstract}
\vfill
{Journal reference: New J. Phys. 19, 122001 (2017)}

\maketitle

\section{Introduction} 

Trapping and cooling nanoscale particles in optical or quasistatic fields at low pressures \cite{li2011,gieseler2012,kiesel2013,asenbaum2013,millen2015,vovrosh2016} offers a
promising route toward the realization of ultra-sensitive force sensors and toward
tests of the quantum superposition principle with massive objects \cite{chang2010,romeroisart2010,ranjit2016}. So far, most experiments  in this nascent field of levitated optomechanics have involved spherical nanoparticles. However, the prospect of enhancing the particle-light interaction by using anisotropic objects \cite{kuhn2015,stickler2016a,hoang2016} motivated several recent experiments demonstrating rotational manipulation and orientational control of aspherical nanoparticles  \cite{kane2010,arita2013,kuhn2015,hoang2016,delord2016,coppock2016,kuhn2017,natcom}.

The rotation of a levitated nanoparticle, governed by its tensor of inertia and by the external torques applied, is in practice always disturbed by ambient gases or fluctuating fields. Individual environmental disruptions can be viewed as random kicks on the angular momentum
vector since they usually take place on short time scales compared to the nanoparticle rotation. A diffusive motion is then expected to occur in the limit of sufficiently small and frequent kicks. This angular
momentum diffusion can be viewed as the weakest conceivable environmental influence
on a rotor, persisting even when the random torque averages to zero.

The purpose of this article is to show how the classical picture of frequent and weak angular momentum kicks can be extended to the quantized description of the rotor and its
interaction with the environment. We derive the general form of the quantum angular
momentum diffusion master equation, show how its diffusion tensors can be related
to the microscopic scattering amplitudes, and demonstrate how the classical diffusion
equation is recovered in the semiclassical limit.

The theory of angular momentum diffusion will be useful to understand
optomechanical setups featuring non-spherical particles. Its quantum 
aspects will become relevant once ro-translational groundstate cooling \cite{stickler2016a,hoang2016,zhong2017} has been achieved, opening the door for quantum experiments
that involve the orientation \cite{shore2015,ma2016,rusconi2016,rusconi2017,delord2017}. Moreover, understanding angular momentum diffusion can be regarded as the first step toward a Markovian quantum theory of rotational friction and thermalization.

We remark that a general microscopic theory of environment-induced decoherence
between different orientation states was presented in \cite{stickler2016b,zhong2016}, and it was also shown
in \cite{stickler2016b}, for the special cases of photon and van der Waals scattering off linear molecules,
that the limit of small anisotropies gives rise to quantum angular momentum diffusion.

We start in Sect.\ \ref{sec:cl} by showing how the classical diffusion equation can be derived from an additive white noise process 
acting on the angular momentum vector. In Sect.\ \ref{sec:der} we then derive a general quantum master equation describing angular momentum diffusion and orientational localization by accounting for the environmental interaction of the quantum rotor in terms of  microscopic, orientation-dependent scattering events. The orientational decoherence and diffusion predicted by the master equation is discussed in Sect.\ \ref{sec:mastereq}, where we also 
evaluate the semiclassical limit of the master equation. 
We then specify our results for the cases of azimuthally symmetric and planar rotors in Sect.\ \ref{sec:symtop}, and   work out explicitly  the associated diffusion constants in Sect.~\ref{sec:micro}, for gas scattering in the Born approximation and for Rayleigh scattering of light.

\section{Classical Angular Momentum Diffusion} \label{sec:cl}

A classical nanoparticle in thermal equilibrium with a homogeneous gas experiences random  kicks through collisions with the gas atoms \cite{gardiner1985,cercignani1988,vankampen1995}. 
The particle's angular momentum vector ${\bf J}$ is then no longer constant in the absence of external potentials, and its dynamics can be described by a Langevin-type stochastic differential equation \cite{vankampen1995}. In the generic case the environment exerts a temporally uncorrelated  random torque $\mathbf{N}(t)$ (with zero mean) whose covariance matrix is characterized by the positive semidefinite tensor $\mathrm{D}(\Omega)$
\begin{equation}\label{eq:Cov}
\left\langle \mathbf{N}(t)\otimes\mathbf{N}(t')\right\rangle=2 \,\mathrm{D}(\Omega)\, \delta(t-t').
\end{equation}
The Langevin equation for ${\bf J}$ then reads as
\begin{equation}
\label{eq:SDE}
\mathrm{d}\mathbf{J}=\sqrt{2\mathrm{D}(\Omega)}\,\mathrm{d}\mathbf{W}_t,
\end{equation} 
with
$\mathrm{d}\mathbf{W}_t$ a vector of standard (uncorrelated) Wiener processes. 
Here, we neglect external torques for convenience.

What turns this equation nontrivial is the fact that the  covariance matrix (\ref{eq:Cov}) naturally depends on the orientation $\Omega$ of the nanoparticle, reflecting that it matters from which direction an impinging gas atom hits the particle surface. 
The orientation $\Omega$, specified e.g.\ by the Euler angles, enters through the orthogonal matrix $\sR(\Omega)$ serving to rotate the particle from an initial orientation to the current one, $\mathrm{D}(\Omega)=\sR(\Omega)\mathrm{D}_0\sR^{\rm T}(\Omega)$. The rotation dynamics, to be solved together with Eq.~(\ref{eq:SDE}), is determined by  \cite{arnold,goldstein}
\begin{equation}
\label{eq:Req}
\mathrm{d}\sR(\Omega)={\rm I}^{-1}(\Omega){\bf J} \times \sR(\Omega) {\rm d}t.
\end{equation} 
Here, ${\rm I}(\Omega) = \sum_{i = 1}^3 I_i {\bf n}_i(\Omega)\otimes{\bf n}_i(\Omega)$ denotes the tensor of inertia, with $I_i$ the moments of inertia and ${\bf n}_i(\Omega)$ the directions of the corresponding principal axes.

As an alternative to the use of stochastic equations, one can solve the deterministic equation for the probability density function $h_t(\Omega,\mathbf{J})$, 
\begin{equation}\label{eq:evol}
\partial_t h_t(\Omega,{\bf J}) = \partial_t^{\rm free} h_t(\Omega,{\bf J}) + \partial_t^{\rm diff} h_t(\Omega,{\bf J}).
\end{equation}
The first term accounts for the free dynamics, the second part describes the effect of the environment. It follows from Eq.~\eqref{eq:SDE}, similar to the derivation of the Fokker-Planck equation \cite{gardiner1985,vankampen1995}, that
\begin{equation}
\label{eq:FokkL}
\partial_t^\mathrm{diff} h_t(\Omega,\mathbf{J})=\nabla_\mathbf{J}\cdot\mathrm{D}(\Omega)\nabla_\mathbf{J}h_t(\Omega,\mathbf{J}).
\end{equation}
Note that the use of canonical momenta instead of $\mathbf{J}$ would render the free evolution in (\ref{eq:evol})  phase space volume preserving, but it would complicate considerably the diffusive part (\ref{eq:FokkL}).

In order to observe that Eq.~\eqref{eq:FokkL} indeed describes angular momentum diffusion, we note that the expectation value of ${\bf J}$ remains constant while its second moment increases linearly with time,
\begin{equation}
\label{eq:classmoment}
\partial_t\left\langle\mathbf{J}\right\rangle=0\qquad \mathrm{and}\qquad \partial_t\left\langle\mathbf{J}\otimes\mathbf{J}\right\rangle=2\, \langle\sD(\Omega)\rangle .
\end{equation}
Hence, we refer to $\sD(\Omega)$ as the angular momentum diffusion tensor in what follows.

The Fokker-Planck equation  of angular momentum diffusion \eqref{eq:FokkL} will be recovered in the semiclassical limit from the quantum master equation discussed in the subsequent section. 

\section{Derivation of the Master Equation} \label{sec:der}

\subsection{Master equation of orientational decoherence}

It is the aim of this section to derive the general form of the quantum angular momentum diffusion  equation by using as a starting point the master equation for the orientational decoherence dynamics of an anisotropic nanoparticle in a homogeneous environment. 
The latter was obtained in \cite{stickler2016b} from the quantum linear Boltzmann equation \cite{vacchini2001test,hornberger2006,hornberger2007,hornberger2008,vacchini2009,smirne2010quantum,hemming2010collisional}. It describes the decay of orientational superposition states due to gas collisions with the nanoparticle surface in terms of the microscopic scattering amplitudes.

The master equation of orientational decoherence is valid if the mean rotation time is much longer than the interaction time so that the scattering amplitude $f(p {\bf n}', p{\bf n}; \Omega)$, describing a single collision between the nanoparticle and a gas atom of incoming and outgoing momentum $p {\bf n}$ and $p {\bf n}'$, is approximately diagonal in the orientation $\Omega$ \cite{stickler2015b,walter2016}. For an asymmetric rotor with state operator $\rho_t$ and  rotational Hamiltonian $\sH$ 
the equation reads as \cite{stickler2016b}
\begin{equation}
\label{eq:master}
\partial_t\rho_t=-\frac{i}{\hbar}\left[\sH,\rho_t\right]+\cL_{\rm g}\rho_t\,,
\end{equation}
with
\begin{eqnarray}
\label{eq:supop1}
\cL_{\rm g}\rho_t&=&\frac{n_\mathrm{g}}{m}\int_0^\infty \mathrm{d}p\,p^3 \mu(p) \int_{S_2}\mathrm{d}^2 \mathbf{n}\mathrm{d}^2\mathbf{n}'\left[f(p {\bf n}', p{\bf n}; \Upomega)\rho_t f^*(p {\bf n}', p{\bf n}; \Upomega) \vphantom{\frac{1}{2}}\right. \nonumber \\
 && \left. -\frac{1}{2}\left\{\vert f(p {\bf n}', p{\bf n}; \Upomega) \vert^2,\rho_t\right\}\right].
\end{eqnarray}
Here, operators are denoted by sans-serif  characters, 
$\mu(p)$ is the (isotropic) momentum distribution of the gas, while $n_{\rm g}$ and $m$ stand for the gas particle density and mass, respectively. 
The master equation \eqref{eq:supop1} is of Lindblad form and thus ensures complete positivity of the evolution $\rho_t$. Note that $ \Upomega$ is an operator rendering the complex, orientation-dependent scattering amplitudes $f(p{\bf n}',p{\bf n};\Upomega)$ operator-valued. 

In order to arrive at the quantum angular momentum diffusion equation we proceed in close analogy to the derivation of momentum diffusion of point particles from the master equation of collisional decoherence \cite{joos1985,breuer2002,vacchini2007relaxation,clark2013diffusive}. 
There one considers the limit of small momentum kicks and expands the Lindblad operators to leading order in the position operator. Likewise, we now consider scattering amplitudes which depend only weakly on the particle orientation, implying that the angular momentum kicks are small.

First, the orientation dependence of the scattering amplitudes can be made explicit by using the isotropy of the scattering process, $f(p {\bf n}',p{\bf n}; \Omega) = f_0[p;{\sR}^{\rm T}(\Omega){\bf n}',\sR^{\rm T}(\Omega){\bf n}]$. Second, the directional dependencies of the scattering amplitude $f_0(p;{\bf n}',{\bf n})$ can be expressed equivalently  in terms of the orientation-independent scattering angle $\cos \theta = {\bf n} \cdot {\bf n}'$ and the orthonormal vectors $({\bf n} \pm {\bf n}')/\vert {\bf n} \pm {\bf n}'\vert$ defining the scattering plane,
\begin{equation}\label{eq:scattamp}
f(p {\bf n}', p {\bf n}; \Omega) = \tilde{f} \left [p,\theta; \sR^{\rm T}(\Omega)\frac{{\bf n} - {\bf n}'}{\vert {\bf n} - {\bf n}' \vert}, \sR^{\rm T}(\Omega) \frac{{\bf n} + {\bf n}'}{\vert {\bf n} + {\bf n}' \vert} \right ].
\end{equation}
The advantage of this is that the orientation now enters only through the scattering plane vectors.

\subsection{Limit of weak dependence on orientation}

We now take the scattering amplitude \eqref{eq:scattamp} to depend 
at most quadratically on the orientation $\Omega$. This limit occurs naturally e.g.\ in Rayleigh scattering of photons if the polarizability tensor is nearly isotropic, or in atom scattering off the nanoparticle surface for near-spherical particles. 

Expanding the directional dependencies in spherical harmonics to second order, the scattering amplitude takes the tensorial form
\begin{eqnarray} \label{eq:scattappr}
\fl
f(p {\bf n}', p{\bf n}; \Omega) & \simeq & f_{\rm sph}(p,\theta) + {\bf A}(p,\theta;\Omega) \cdot \frac{{\bf n} - {\bf n}'}{\vert {\bf n} - {\bf n}' \vert}  \nonumber \\
 && + \frac{{\bf n} - {\bf n}'}{\vert {\bf n} - {\bf n}' \vert}  \cdot \sB(p,\theta;\Omega) \frac{{\bf n} - {\bf n}'}{\vert {\bf n} - {\bf n}' \vert}+ \frac{{\bf n} + {\bf n}'}{\vert {\bf n} + {\bf n}' \vert} \cdot \sC(p,\theta;\Omega) \frac{{\bf n} + {\bf n}'}{\vert {\bf n} + {\bf n}' \vert},
\end{eqnarray}
with the scalar $f_{\rm sph}(p,\theta)$, the vector ${\bf A}(p,\theta;\Omega)$ and the symmetric tensors $\sB(p,\theta;\Omega)$, $\sC(p,\theta;\Omega)$ as coefficients. Here we used that time reversal invariance $f(p{\bf n}',p{\bf n};\Omega) = f(-p{\bf n},-p{\bf n}';\Omega)$ \cite{taylor}, removes all terms linear in ${\bf n}+{\bf n}'$.
Note that the scattering amplitude  $f_{\rm sph}$ accounts for spherically symmetric scattering, and that the vector ${\bf A}(p,\theta;\Omega) = \sR(\Omega) {\bf A}_0(p,\theta)$ vanishes if the scattering amplitude is inversion symmetric, $f(- p{\bf n}',- p{\bf n};\Omega) = f(p{\bf n}',p{\bf n};\Omega)$. In the latter case the scattering amplitude depends quadratically on the nanoparticle's orientation, as described by the symmetric tensors $\sB(p,\theta;\Omega) = \sR(\Omega) \sB_0(p,\theta) \sR^{\rm T}(\Omega)$ and $\sC(p,\theta;\Omega) = \sR(\Omega) \sC_0(p,\theta) \sR^{\rm T}(\Omega)$.

Inserting the scattering amplitude \eqref{eq:scattappr} into Eq.\ \eqref{eq:supop1} yields two terms, one depending linearly and one quadratically on the particle's orientation,
\begin{eqnarray} \label{eq:qmaster}
\partial_t \rho_t = -\frac{i}{\hbar}\left[\sH,\rho_t\right]+\cL_{1}\rho_t+\cL_{2}\rho_t,
\end{eqnarray}
with
\numparts
\begin{eqnarray} \label{eq:sing}
\cL_{1} \rho_t & = & \frac{2 \pi n_\mathrm{g}}{m}\int_0^\infty \mathrm{d}p\,p^3 \mu(p) \int_{0}^\pi \dint{\theta} \sin \theta \int_{S_2}\mathrm{d}^2 \mathbf{n} \nonumber \\
&& \left[\left [\mathbf{n}\cdot\mathbf{A}(p,\theta;\Upomega) \right ] \rho_t \left [\mathbf{n}\cdot\mathbf{A}^*(p,\theta;\Upomega) \right ] -\frac{1}{2}\left\{ \left \vert \mathbf{n}\cdot\mathbf{A}(p,\theta;\Upomega) \right \vert^2,\rho_t\right\}\right],\nonumber 
\\&&
\end{eqnarray}
and
\begin{eqnarray} \label{eq:quad}
\cL_{2} \rho_t & = & \frac{2 \pi n_\mathrm{g}}{m}\int_0^\infty \mathrm{d}p\,p^3 \mu(p) \int_0^\pi \dint{\theta} \sin\theta \int_{S_2}\mathrm{d}^2 \mathbf{n} \nonumber \\
 && \left[\left [\mathbf{n}\cdot\sB(p,\theta;\Upomega)\mathbf{n} \right ] \rho_t \left [\mathbf{n}\cdot\sB^*(p,\theta;\Upomega)\mathbf{n} \right ] -\frac{1}{2}\left\{\left \vert\mathbf{n}\cdot\sB(p,\theta;\Upomega)\mathbf{n} \right \vert^2,\rho_t\right\}\right] \nonumber \\
 && + \frac{2 \pi n_\mathrm{g}}{m}\int_0^\infty \mathrm{d}p\,p^3 \mu(p) \int_0^\pi \dint{\theta} \sin\theta \int_{S_2}\mathrm{d}^2 \mathbf{n} \nonumber \\
 && \left[\left [\mathbf{n}\cdot\sC(p,\theta;\Upomega)\mathbf{n} \right ] \rho_t \left [\mathbf{n}\cdot\sC^*(p,\theta;\Upomega)\mathbf{n} \right ] -\frac{1}{2}\left\{\left \vert\mathbf{n}\cdot\sC(p,\theta;\Upomega)\mathbf{n} \right \vert^2,\rho_t\right\}\right].
\nonumber 
\\&&
\end{eqnarray}
\endnumparts

This concludes the general derivation of the  angular momentum diffusion equation for isotropic environments. (Anisotropic environments can be treated analogously by taking the momentum distribution in (\ref{eq:supop1}) to depend on direction $\mathbf{n}$.)
Microscopic expressions for the expansion coefficients will be provided in Sec.\ \ref{sec:micro} for particle scattering in the Born approximation and incoherent photon scattering.
In order to discuss the properties of the master equation (\ref{eq:qmaster}) we now reduce it to its most elementary form.

\subsection{Generic master equation of angular momentum diffusion}

In a first step, we note that the momentum integrals in Eq.~\eqref{eq:sing} and \eqref{eq:quad} can be omitted since they amount to a sum of superoperators of equivalent form. This is physically justified because the master equation \eqref{eq:qmaster} describes orientational decoherence also in the case that the momentum distribution $\mu(p)$ is sharply peaked. Since the same argument applies to the integration over the scattering angle $\theta$ we will neglect also the dependence on the scattering angle in the following discussion. Second, we omit one of the two terms contributing to the quadratic superoperator \eqref{eq:quad} because they are of the same form and therefore describe the same physical effects.

Finally, we assume for simplicity that the vector ${\bf A}(\Omega)$ and the tensor $\sB(\Omega)$ are real. This is not a severe restriction since the sole effect of complex Lindblad operators on orientational coherences  $\matel{\Omega}{\rho_t}{\Omega'}$ is to induce an additional  phase oscillation, whose frequency is proportional to ${\rm Im}[f(p {\bf n}', p {\bf n},\Omega)f^*(p {\bf n}', p {\bf n},\Omega')]$ \cite{stickler2016b} while the decoherence rates are unaffected. The phase oscillation vanishes if the Lindblad operators ${\bf A}(\Upomega)$ and $\sB(\Upomega)$ are either hermitian or anti-hermitian.

	These simplifications yields the elementary superoperators 
\numparts
\begin{eqnarray} 
\label{eq:suplin}
\cL_1\rho_t=& \int_{S_2}\frac{\dint{^2\mathbf{n}}}{4\pi}\left[ \left [\mathbf{n}\cdot\mathbf{A}(\Upomega) \right ]\rho_t \left [\mathbf{n}\cdot\mathbf{A}(\Upomega) \right ]-\frac{1}{2}\left\{\left[\mathbf{n}\cdot\mathbf{A(\Upomega)}\right]^2,\rho_t\right\}\right],\\
\label{eq:supquad}
\cL_\mathrm{2}\rho_t=& \int_{S_2}\frac{\dint{^2\mathbf{n}}}{4\pi}\left[ \left [\mathbf{n}\cdot\sB(\Upomega)\mathbf{n} \right ]\rho_t \left [\mathbf{n}\cdot\sB(\Upomega)\mathbf{n} \right]-\frac{1}{2}\left\{\left[\mathbf{n}\cdot\sB(\Upomega)\mathbf{n}\right]^2,\rho_t\right\}\right].
\end{eqnarray}
\endnumparts
The orientation operator $\Upomega$ enters through the real vector $\mathbf{A}(\Omega)=\sR(\Omega)\mathbf{A}_0$ and the symmetric tensor $\sB(\Omega)=\sR(\Omega)\sB_0\sR^{\rm T}(\Omega)$, which describe the influence of the environment. Here we absorbed all prefactors into the Lindblad operators for convenience.

Carrying out the integrations and exploiting the spectral decomposition of $\sB(\Upomega)$ 
Eqs.\ (\ref{eq:suplin}) and (\ref{eq:supquad}) can be written more compactly as
\numparts
\begin{eqnarray}
\label{eq:suplin2}
\cL_1\rho_t=& \frac{1}{3} \left ( \mathbf{A}(\Upomega) \cdot \rho_t \mathbf{A}(\Upomega) -A^2  \rho_t \right ),\\
\label{eq:supquad2}
\cL_\mathrm{2}\rho_t=& \frac{2}{15}{\rm Tr}\left[ \sB(\Upomega)\rho_t \sB(\Upomega)-\frac{1}{2}\left\{\sB^2(\Upomega),\rho_t\right\}\right].
\end{eqnarray}
\endnumparts
Here ${\rm Tr}(\cdot)$ stands for the matrix trace, not to be confused with the operator trace  acting on the Hilbert space of orientation states. Moreover, $A=| \mathbf{A}(\Omega)|$ is the (orientation-independent) length of the vector $ \mathbf{A}(\Omega)$.

\section{Quantum Angular Momentum Diffusion} \label{sec:mastereq}

We now prove that the master equation \eqref{eq:qmaster} 
describes orientational decoherence, gives rise to angular momentum diffusion, and turns in the semiclassical limit into the expected classical angular momentum diffusion equation.

\subsection{Orientational decoherence}

An important property of the angular momentum diffusion superoperators \eqref{eq:suplin} and \eqref{eq:supquad} is that they jointly serve to completely localize the orientation of the nanoparticle, in the sense that they reduce all orientational coherences. Equations  \eqref{eq:suplin} and \eqref{eq:supquad} give rise to an exponential decay of the orientational coherences of the state operator since the Lindblad operators are diagonal in the orientation,
\begin{eqnarray}
\left\langle \Omega\right\vert\cL_{1,2} \rho_t\left\vert\Omega'\right\rangle=-F_{1,2}(\Omega,\Omega')\left\langle\Omega\right\vert\rho_t\left\vert\Omega'\right\rangle.
\end{eqnarray}
We proceed to prove that the total localization rate $F_{1}(\Omega,\Omega')+F_{2}(\Omega,\Omega')$ is indeed positive for $\Omega\neq\Omega'$ and  vanishes for $\Omega=\Omega'$. Moreover, it depends only on the relative orientation between $\Omega$ and $\Omega'$. 

As a first step, one can show that the localization rates take the explicit form
\numparts
\begin{eqnarray}
\label{eq:difflocratelin}
F_1(\Omega,\Omega')=\frac{2 D^{(1)}}{\hbar^2}\left[1-\mathbf{a}(\Omega)\cdot\mathbf{a}(\Omega')\right]\\
\label{eq:difflocratequad}
F_\mathrm{2}(\Omega,\Omega')=\frac{1}{2\hbar^2}\sum_{i=1}^3{\left(\sum_{j=1}^3 D_j^{(2)}-2D_i^{(2)}\right)\left\vert\mathbf{b}_i(\Omega)\times\mathbf{b}_i(\Omega')\right\vert^2},
\end{eqnarray}
\endnumparts
where we set ${\bf A}(\Omega)=A {\bf a}(\Omega) $
and $\sB(\Omega)=\sum_{i=1}^3B_i  {\bf b}_i(\Omega)\otimes{\bf b}_i(\Omega)$
with unit vectors ${\bf a}(\Omega)$ and ${\bf b}_i(\Omega)$. Moreover, 
we defined the diffusion constants
\begin{equation}
\label{eq:diffconst}
D^{(1)}=\frac{\hbar^2}{6}A^2 \quad \mathrm{and} \quad
D_i^{(2)}=\frac{2\hbar^2}{15}(B_j-B_k)^2,
\end{equation}
with $(i, j, k)$ cyclic permutations of $(1,2,3)$. The reasons for choosing this definition  will become obvious in Sec.~\ref{subsec:angmomdiff}.

Note that the expressions (\ref{eq:difflocratelin}) and (\ref{eq:difflocratequad}) already imply that the localization rates depend only on the angle of rotation between $\Omega$ and $\Omega'$ since ${\bf a}(\Omega)=\sR(\Omega){\bf a}(0)$ and ${\bf b}_i(\Omega)=\sR(\Omega){\bf b}_i(0)$. Moreover, it is evident from (\ref{eq:difflocratelin}) that $F_1(\Omega,\Omega') \geq 0 $. However, the equality holds for all ${\bf a}(\Omega) = {\bf a}(\Omega')$ meaning  that (\ref{eq:difflocratelin}) vanishes for superpositions between orientation states related by a rotation around ${\bf a}(\Omega)$. Several independent superoperators of form (\ref{eq:suplin}) would therefore be required to achieve complete localization. Alternatively, the addition of Eq.\ (\ref{eq:supquad}) can induce complete localization.

In order to show that $F_2(\Omega,\Omega') > 0$ for $\Omega\neq \Omega'$ we note that at most one of the coefficients $f_i = \sum_{i = 1}^3 D_j^{(2)}-2D_i^{(2)}$ can be negative,  say $f_3 < 0$, and $\vert f_3 \vert < f_{1,2}$, as follows from the positivity of the diffusion coefficients (\ref{eq:diffconst}). A direct calculation yields
\begin{eqnarray}
\label{eq:F2abs}
\fl F_2(\Omega,\Omega') & = & \sum_{i = 1}^3 f_i \vert {\bf b}_i(\Omega) \times {\bf b}_i(\Omega') \vert^2 \nonumber \\
& \geq& \vert f_3 \vert \left ( \vert {\bf b}_1(\Omega) \times {\bf b}_1(\Omega') \vert^2 + \vert {\bf b}_2(\Omega) \times {\bf b}_2(\Omega') \vert^2 - \vert {\bf b}_3(\Omega) \times {\bf b}_3(\Omega') \vert^2 \right ) \nonumber \\
& = & \vert f_3 \vert \left (2 - \vert {\bf b}_2(\Omega) \times {\bf b}_1(\Omega') \vert^2 - \vert {\bf b}_1(\Omega) \times {\bf b}_2(\Omega') \vert^2 \right ),
\end{eqnarray}
where we used that any orthonormal set ${\bf b}_i(\Omega)$ and any unit vector ${\bf c}$ satisfies 
\begin{equation}
\sum_{i = 1}^3 \vert {\bf b}_i(\Omega) \times {\bf c} \vert^2 = 2.
\end{equation}
The expression (\ref{eq:F2abs}) shows that the localization rate $F_2$ is manifestly positive unless ${\bf b}_1(\Omega)\bot {\bf b}_2(\Omega')$ and ${\bf b}_2(\Omega)\bot {\bf b}_1(\Omega')$. The only pairs of orientations whose coherences do not get localized are therefore related by a $\pi$-rotation around ${\bf b}_3(\Omega)$. In general, {\it i.e.}\ for ${\bf a}(\Omega) \neq {\bf b}_3(\Omega)$, this is remedied by the decoherence due to $\cL_1 \rho$.

\subsection{Angular momentum diffusion} \label{subsec:angmomdiff}

Each of the superoperators $\cL_1$ and $\cL_2$ induces quantum angular momentum diffusion. We demonstrate this by calculating explicitly the time evolution of the expectation value of the angular momentum operator $\boldsymbol{\sJ}$ and of its tensor of second moments $\boldsymbol{\sJ}\otimes\boldsymbol{\sJ}$. As in the classical case (\ref{eq:classmoment}), we expect that in the absence of a potential
\begin{equation}\label{eq:qfm}
\partial_t\left\langle\boldsymbol{\sJ}\right\rangle = 0, \quad {\rm and } 
\quad \partial_t\left\langle\boldsymbol{\sJ} \otimes \boldsymbol{\sJ} \right\rangle = 2 \left \langle \sD(\Upomega)\right \rangle  ,
\end{equation}
where the angular brackets now stand for the quantum expectation value.
Moreover, we will see in the subsequent section that the semiclassical limit of $\cL_1\rho$ and $\cL_2\rho$ yields the classical diffusion equation (\ref{eq:FokkL}). It is therefore natural to refer to the dynamics induced by $\cL_1$ and $\cL_2$  as quantum angular momentum diffusion.

In order to derive Eqs.\ (\ref{eq:qfm}) we choose an explicit parametrization of the orientation $\Omega$ in terms of Euler angles $(\alpha,\beta,\gamma)$ in the $z$-$y'$-$z''$-convention \cite{edmonds1996,brink2002,timothesis}. In particular, an $\alpha$-rotation around the ${\bf e}_z$ axis is followed by a $\beta$-rotation around the nodal line ${\bf e}_\nu(\alpha) = -{\bf e}_x  \sin \alpha + {\bf e}_y\cos \alpha $, and a final $\gamma$-rotation around the the body-fixed axis ${\bf n}_3(\beta,\alpha)$. Note that ${\bf n}_3(\Omega) = \sR(\Omega){\bf e}_z$.

The space-fixed components of the angular momentum operator can be expressed as \cite{edmonds1996}
\numparts
\begin{eqnarray} 
\sJ_x&=-\left(\frac{\cot\upbeta}{2}\left\{\pop_\alpha,\cos\upalpha\right\}+\sin\upalpha \,\pop_\beta-\frac{\cos\upalpha}{\sin\upbeta}\pop_\gamma\right),\\
\sJ_y&=-\left(\frac{\cot\upbeta}{2}\left\{\pop_\alpha,\sin\upalpha\right\}-\cos\upalpha \,\pop_\beta-\frac{\sin\upalpha}{\sin\upbeta}\pop_\gamma\right),\\
\sJ_z&=\pop_\alpha.
\end{eqnarray}
\endnumparts
with the momentum operators canonically conjugate to the Euler angle operators $\upalpha, \upbeta, \upgamma$ given by
\begin{eqnarray}
\pop_\alpha = \frac{\hbar}{i} \frac{\partial}{\partial \alpha}, \qquad \pop_\beta = \frac{\hbar}{i} \left ( \frac{\partial}{\partial \beta} + \frac{1}{2} \cot \beta \right ), \qquad \pop_\gamma = \frac{\hbar}{i} \frac{\partial}{\partial \gamma}
\end{eqnarray}
To evaluate the expectation values in (\ref{eq:qfm}) one makes use of the commutation relations between a function of the orientation  and the canonical momentum operators  
\begin{equation}
\left[g(\upalpha, \upbeta, \upgamma),\pop_\lambda\right]=i\hbar\partial_{\lambda}g(\upalpha, \upbeta, \upgamma).
\end{equation}
Exploiting the fact that the free Hamiltonian commutes with the angular momentum one readily obtains $\partial_t\left\langle\boldsymbol{\sJ}\right\rangle = 0$. Similarly, a lengthy but straightforward calculation yields 
\begin{equation}\label{eq:qfm2}
\quad \partial_t\left\langle\boldsymbol{\sJ} \otimes \boldsymbol{\sJ} \right\rangle = 2 \left \langle \sD^{(1)}(\Upomega) + \sD^{(2)}(\Upomega) \right \rangle,
\end{equation}
with the  tensors
\numparts
\begin{eqnarray}\label{eq:Diften1}
\sD^{(1)}(\Omega) & = & D^{(1)} \left [ \mathds{1} - {\bf a}(\Omega) \otimes {\bf a}(\Omega) \right ]  \\ 
\label{eq:Diften2}
\sD^{(2)}(\Omega) & = & \sum_{i = 1}^3 D^{(2)}_i {\bf b}_i(\Omega) \otimes {\bf b}_i(\Omega).
\end{eqnarray}
\endnumparts
This implies  that the angular momentum variance increases linearly with time since $\partial_t\left\langle\boldsymbol{\sJ}^2\right\rangle = 4 D^{(1)} + 2 \sum_{i=1}^3D_i^{(2)}.$ 
The tensors (\ref{eq:Diften1}), (\ref{eq:Diften2}) can thus be identified as the angular momentum diffusion tensors. Their eigenvalues, introduced in Eq.\ (\ref{eq:diffconst}), are the corresponding diffusion coefficients.

\subsection{Semiclassical limit} \label{subsec:scl}

To obtain the semiclassical limit of the quantum angular momentum diffusion equation  \eqref{eq:qmaster}  we express it in quantum phase space using the Wigner function for the orientation state and retain  only the leading order in $\hbar$. 
Since the space of orientations is compact the associated momenta are discrete. Using the Euler angles $\Omega=(\alpha, \beta, \gamma)$ as configuration space coordinates, the Wigner function is given by \cite{fischer2013,gneiting2013} 
\begin{eqnarray} \label{eq:rotwig}
\fl w_t(\Omega, {\bf m}_\Omega) = & \frac{1}{4 \pi^3} \int_{-\pi}^\pi \mathrm{d} \alpha' \int_{-\pi/2}^{\pi/2} \mathrm{d} \beta' \int_{-\pi}^\pi \mathrm{d} \gamma' \sqrt{\sin \beta_+ \sin \beta_-} e^{i (m_\alpha \alpha'+ 2 m_\beta \beta'+ m_\gamma \gamma')} \nonumber \\
 & \times \left \langle \alpha_- \beta_- \gamma_- \vert \rho_t \vert \alpha_+ \beta_+ \gamma_+ \right \rangle,
\end{eqnarray}
where ${\bf m}_\Omega = (m_\alpha,m_\beta,m_\gamma) \in \ZZ^3$ labels the discrete spectrum of the canonical momentum operators $(\pop_\alpha,\pop_\beta,\pop_\gamma)$. In addition, we abbreviated $\alpha_\pm=(\alpha\pm\alpha'/2)\,\mathrm{mod}\,2\pi$, likewise for $\gamma_\pm$, and $\beta_\pm = (\beta\pm\beta'/2)\,\mathrm{mod}\,\pi$.

With this definition the master equation \eqref{eq:qmaster} takes the form
\begin{equation}
\label{eq:Weyltime}
\partial_t w_t(\Omega,\mathbf{m}_\Omega)= -\frac{i}{\hbar}\left ( h \star w_t - w_t \star h \right )(\Omega,\mathbf{m}_\Omega) - \frac{1}{4 \pi^3} \cQ\left[w_t\right](\Omega,\mathbf{m}_\Omega),
\end{equation}
where $h(\Omega,\mathbf{m}_\Omega)$ is the phase space transform of the Hamiltonian $\sH$ and the symbol $ \star $ denotes the star product defined in \cite{gneiting2013}. The first term accounts for the unitary rotational quantum dynamics of the rigid top and reduces to the Poisson bracket in the semiclassical limit \cite{fischer2013,gneiting2013}. 

The diffusive motion is described by
\begin{eqnarray}
\label{eq:Weylsup}
\fl\cQ\left[w_t\right](\Omega,\mathbf{m}_\Omega)= & \sum_{\mathbf{m}_\Omega'\in\ZZ^3}\int_{-\pi}^\pi \mathrm{d}\alpha' \int_{-\pi/2}^{\pi/2} \mathrm{d} \beta' \int_{-\pi}^\pi \mathrm{d} \gamma'\mathrm{e}^{i (m_\alpha-m'_\alpha)\alpha'}\mathrm{e}^{i2(m_\beta-m'_\beta)\beta'}\mathrm{e}^{i(m_\gamma-m'_\gamma)\gamma'} \nonumber \\
&\times\left[F_1(\Omega_-,\Omega_+) + F_2(\Omega_-,\Omega_+)\right]w_t(\Omega,\mathbf{m}'_\Omega).
\end{eqnarray}
The semiclassical limit of (\ref{eq:Weylsup}) can now be calculated in three steps. First, we consider the state of a sufficiently massive particle implying that its Wigner function varies weakly with the angular momentum quantum numbers ${\bf m}_\Omega$. The latter can thus be approximated by the continuous angular momenta $(p_\alpha,p_\beta,p_\gamma) \simeq (\hbar m_\alpha, 2 \hbar m_\beta, \hbar m_\gamma)$ and accordingly the sum over angular momentum quantum numbers is replaced by an integral. The following  steps consist of expanding the integrand in Eq.\ (\ref{eq:Weylsup}) to leading order in $\hbar$ and then rewriting the resulting equation in terms of the angular momentum vector in order to compare with the classical diffusion equation (\ref{eq:FokkL}).

By substituting $\xi_\alpha = \alpha'/\hbar$ in (\ref{eq:Weylsup}) we ensure that $\hbar$ appears only in the arguments of $F_{1,2}$ in the form $\alpha \pm \hbar \xi_\alpha/2$, and accordingly for $\beta'$ and $\gamma'$. Retaining $F_{1,2}$ up to second order in $\hbar$ requires expanding the vectors ${\bf a}(\Omega \pm \hbar \xi_\Omega/2)$ up to second order in $\hbar$ and ${\bf b}_i(\Omega \pm \hbar \xi_\Omega/2)$ to first order, see Eqs.\ (\ref{eq:difflocratelin}) and  (\ref{eq:difflocratequad}). Here it is pertinent to expand in a coordinate-independent way rather than componentwise. For example,
\begin{eqnarray}
\fl
{\bf a}(\Omega \pm \hbar \xi_\Omega/2) & = & {\bf a} \pm \frac{\hbar \xi_\alpha}{2} {\bf e}_z \times {\bf a} \pm \frac{\hbar \xi_\beta}{2} {\bf e}_\nu \times {\bf a} \pm \frac{\hbar \xi_\gamma}{2} {\bf n}_3 \times {\bf a} \nonumber \\
&& + \frac{\hbar^2 \xi_\alpha^2}{8}{\bf e}_z \times\left ( {\bf e}_z \times {\bf a} \right ) + \frac{\hbar^2 \xi_\beta^2}{8}{\bf e}_\nu \times\left ( {\bf e}_\nu \times {\bf a} \right ) + \frac{\hbar^2 \xi_\gamma^2}{8}{\bf n}_3 \times\left ( {\bf n}_3 \times {\bf a} \right ) \nonumber \\
&& + \frac{\hbar^2 \xi_\alpha \xi_\beta}{8} \left ( {\bf e}_z \otimes {\bf e}_\nu + {\bf e}_\nu \otimes {\bf e}_z \right ){\bf a} + \frac{\hbar^2 \xi_\beta \xi_\gamma}{8} \left ( {\bf n}_3 \otimes {\bf e}_\nu + {\bf e}_\nu \otimes {\bf n}_3 \right ){\bf a} \nonumber \\
&& + \frac{\hbar^2 \xi_\alpha \xi_\gamma}{8} \left ( {\bf}{\bf e}_z \otimes {\bf n}_3 + {\bf n}_3 \otimes {\bf e}_z \right ){\bf a} - \frac{\hbar^2 \xi_\alpha \xi_\gamma}{4} ({\bf n}_3 \cdot {\bf e}_z) {\bf a},
\end{eqnarray}
where we dropped the $\Omega$-dependence of ${\bf a}$, ${\bf e}_\nu$ and ${\bf n}_3$. In the limit of small $\hbar$ the integration boundaries of $\xi_\Omega$ tend to infinity, which allows one to carry out all integrals and rewrite Eq.\ (\ref{eq:Weylsup}) in terms of $p_\Omega$-derivatives.

Finally, by transforming the Wigner distribution function to the space of orientation and angular momenta $h_t(\Omega,{\bf J})$, one obtains after a lengthy but straightforward calculation
\begin{equation}
\partial_t^\mathrm{diff} h_t(\Omega,\mathbf{J})=\nabla_\mathbf{J}\cdot\left [ \mathrm{D}^{(1)}(\Omega) + \mathrm{D}^{(2)}(\Omega) \right ]\nabla_\mathbf{J}h_t(\Omega,\mathbf{J}) + \cO(\hbar),
\end{equation}
with the two diffusion tensors (\ref{eq:Diften1}) and (\ref{eq:Diften2}). Thus the classical angular momentum diffusion equation (\ref{eq:FokkL}) emerges naturally as the semiclassical limit of the two superoperators \eqref{eq:suplin} and \eqref{eq:supquad}.

\section{Symmetric Nanoparticles} \label{sec:symtop}

We now specify the master equation of quantum angular momentum diffusion for symmetric particles. This simplifies the equation and facilitates its use in practical applications. We start with azimuthally symmetric rotors and then turn to the planar rotor, for which the master equation can be solved explicitly.
 
\subsection{Symmetric and linear rigid rotor} 

In the case of perfect azimuthal symmetry, any physical interaction between the nanoparticle and its environment depends at most on the direction of the nanoparticle's symmetry axis ${\bf m}(\Omega)$. Prominent examples include the anisotropic van der Waals interaction \cite{Stonebook} or Rayleigh-Gans photon scattering \cite{dehulst,schiffer79}. Accounting for this symmetry,  we assume the vector ${\bf a}(\Omega)$ in (\ref{eq:suplin}) to point along ${\bf m}(\Omega)$ and the tensor $\sB(\Omega)$ in (\ref{eq:supquad}) to be $\sB(\Omega) = \mathds{1}B_\perp+(B_\parallel-B_\perp)\mathbf{m}(\Omega)\otimes\mathbf{m}(\Omega)$.

Inserting ${\bf A}(\Omega) = A {\bf m}(\Omega)$ and $\sB(\Omega)$ into the superoperators \eqref{eq:suplin} and \eqref{eq:supquad} yields
\numparts\begin{eqnarray}
\label{eq:supoplrlin}
\fl\cL_1\rho_t=\frac{6 D^{(1)}}{\hbar^2}\int_{S_2}\frac{\mathrm{d}^2\mathbf{n}}{4\pi}\left[[\mathbf{n}\cdot \mathbf{m}(\Omega)]\rho_t[\mathbf{n}\cdot \mathbf{m}(\Omega)]-\frac{1}{2}\left\{[\mathbf{n}\cdot \mathbf{m}(\Omega)]^2,\rho_t\right\}\right]\\
\label{eq:supoplrquad}
\fl\cL_2\rho_t=\frac{15 D^{(2)}}{2 \hbar^2}\int_{S_2}\frac{\dint{^2\mathbf{n}}}{4\pi}\left[\left[\mathbf{n}\cdot\mathbf{m}(\Omega)\right]^2\rho_t\left[\mathbf{n}\cdot\mathbf{m}(\Omega)\right]^2-\frac{1}{2}\left\{\left[\mathbf{n}\cdot\mathbf{m}(\Omega)\right]^4,\rho_t\right\}\right].
\end{eqnarray}\endnumparts
where we used the diffusion constants \eqref{eq:diffconst}
\begin{equation}\label{eq:diffconstsym}
D^{(1)} = \frac{\gamma_1 \hbar^2 A^2}{6} \quad \textrm{and} \quad D^{(2)} = \frac{2\gamma_2 \hbar^2}{15}\left(B_\perp-B_\parallel\right)^2.
\end{equation}
An equation of the form \eqref{eq:supoplrquad} has already been derived in Ref. \cite{stickler2016b} for the special cases of photon scattering in the generalized Rayleigh-Gans approximation and atom scattering off the anisotropic homogeneous  dipole-induced dipole potential.

From the discussion in the previous section it follows that the second moment of the angular momentum vector increases linearly with time as $\partial_t \langle \boldsymbol{\sJ}^2 \rangle = 4(D^{(1)} + D^{(2)})$. The resulting orientational localization rates can be written compactly as
\numparts
\begin{eqnarray}
F_1(\Omega,\Omega')=\frac{2 D^{(1)}}{\hbar^2}\left[1-\mathbf{m}(\Omega)\cdot\mathbf{m}(\Omega')\right],\\
F_2(\Omega,\Omega')=\frac{D^{(2)}}{\hbar^2}\left\vert \mathbf{m}(\Omega)\times\mathbf{m}(\Omega')\right\vert^2.
\end{eqnarray}
\endnumparts
In the semiclassical limit, one obtains the phase space angular momentum diffusion equation
\begin{equation}
\label{eq:diffsymm}
\partial_t^\mathrm{diff} w_t(\Omega,p_\Omega)=(D^{(1)} + D^{(2)}) (\sin^2\beta\partial^2_{p_\alpha}+\partial^2_{p_\beta})w_t(\Omega,p_\Omega) + O(\hbar).
\end{equation}
The Langevin equation \eqref{eq:SDE} gives rise to the same diffusion equation in the limit that no rotations around the symmetry axis ${\bf m}(\Omega)$ are considered.

\subsection{Planar rigid rotor}

The diffusion master equation simplifies even further if the motion is restricted to planar rotations parametrized by the single degree of freedom $\alpha \in [0,2\pi)$. Replacing in Eqs.~\eqref{eq:supoplrlin} and \eqref{eq:supoplrquad} ${\bf m}(\Omega)$ by ${\bf e}_x\cos \alpha  +  {\bf e}_y \sin \alpha $ and carrying out the integration over the polar angle yields
\numparts\begin{eqnarray} 
\label{eq:supoppr}
\fl\cL_1\rho_t=\frac{2 D^{(1)}}{\pi\hbar^2}\int_0^{2\pi}\dint{\varphi}\left[\cos(\upalpha-\varphi)\rho_t\cos(\upalpha-\varphi)-\frac{1}{2}\left\{\cos^2(\upalpha-\varphi),\rho_t\right\}\right],\\
\label{eq:supopprq}
\fl\cL_2\rho_t=\frac{2 D^{(2)}}{\pi\hbar^2}\int_0^{2\pi}\dint{\varphi}\left[\cos^2(\upalpha-\varphi)\rho_t\cos^2(\upalpha-\varphi)-\frac{1}{2}\left\{\cos^4(\upalpha-\varphi),\rho_t\right\}\right].
\end{eqnarray}\endnumparts
The localization rates are
\numparts\begin{eqnarray} 
F_1(\alpha,\alpha')=\frac{4 D^{(1)}}{\hbar^2}\sin^2\left(\frac{\alpha-\alpha'}{2}\right)\\
F_2(\alpha,\alpha')=\frac{D^{(2)}}{\hbar^2}\sin^2(\alpha-\alpha').
\end{eqnarray}\endnumparts
The semiclassical limit of the superoperators \eqref{eq:supoppr} and \eqref{eq:supopprq} gives the diffusion equation \eqref{eq:diffsymm} with $\beta=\pi/2$ and without the $p_\beta$ derivative. Note that Eq.~\eqref{eq:supopprq} appears already in Ref. \cite{schrinski2017}, where it is obtained from the collapse theory of continuous spontaneous localization (CSL).

In order to solve the master equation (\ref{eq:qmaster}) with the rotational Hamiltonian $\sH_\mathrm{free}=\pop_{\alpha}^2/2I$ and superoperators \eqref{eq:supoppr} and \eqref{eq:supopprq} we express it in quantum phase space by using the 1D version of the Wigner function (\ref{eq:rotwig}),
\begin{eqnarray}
\label{eq:wignerpl}
\fl \partial_t w_t(\alpha,m_\alpha) + & \frac{\hbar m_\alpha}{I}\partial_\alpha w_t(\alpha,m_\alpha) = \frac{D^{(1)}}{\hbar^2} \left [w_t(\alpha,m_\alpha+1)-2w_t(\alpha,m_\alpha)+w_t(\alpha,m_\alpha-1) \right ] \nonumber \\
 & + \frac{D^{(2)}}{(2 \hbar)^2} \left[w_t(\alpha,m_\alpha+2)-2w_t(\alpha,m_\alpha)+w_t(\alpha,m_\alpha-2) \right ].
\end{eqnarray}
The left hand side of this equation describes the classical shearing dynamics of the rotor while diffusion enters on the right hand side in the form of discretized angular momentum derivatives. The fact that next-to-nearest angular momenta are coupled in the last term reflects the $\pi$-symmetry of the localization rate $F_2$.

\begin{figure}
\label{fig:planar}
\centering
\includegraphics[width=\textwidth]{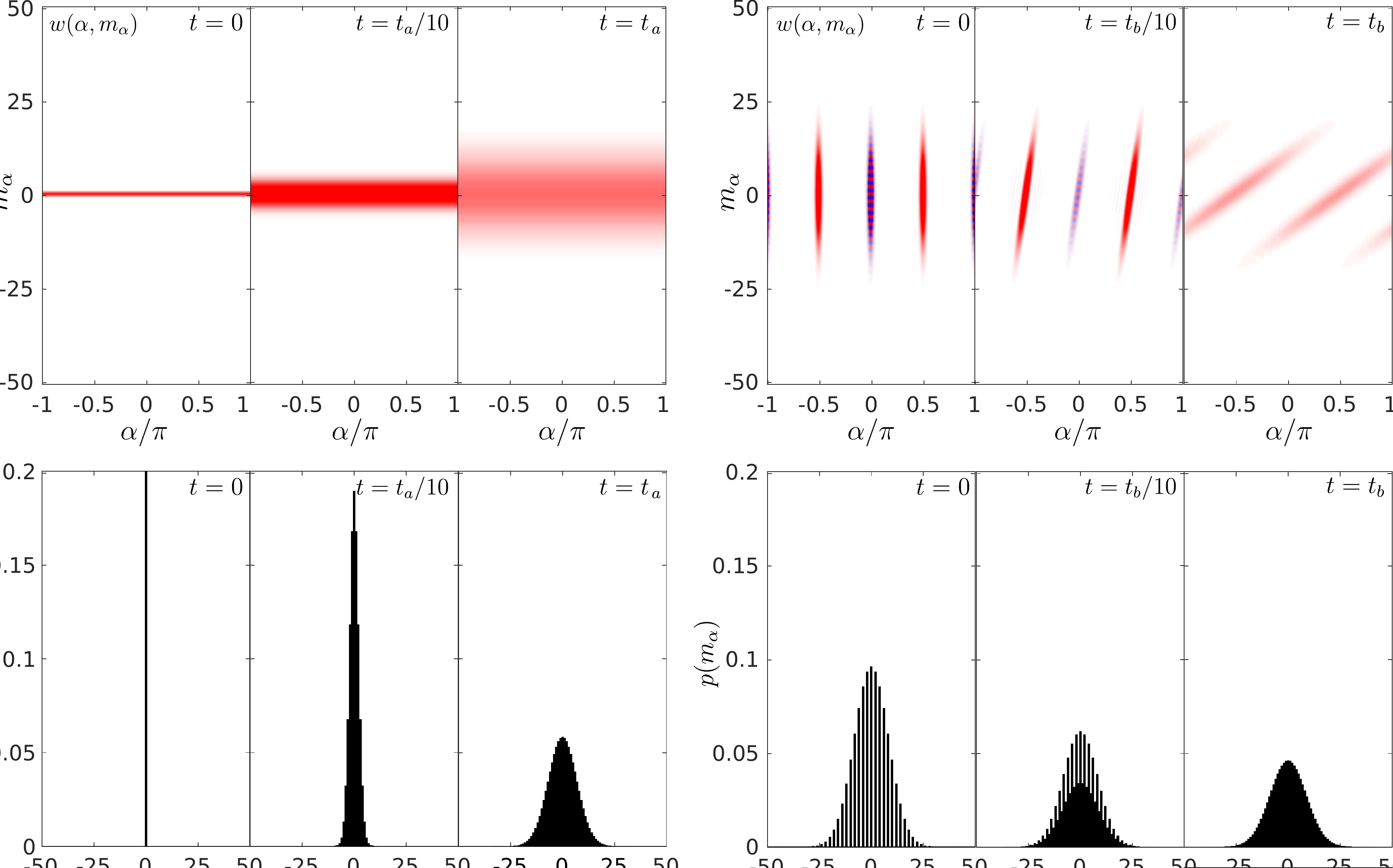}
\caption{(a) Time evolution of the rotational ground state  in presence of angular momentum diffusion. We show a density plot of the Wigner function $w_t(\alpha,m_\alpha)$ and the momentum distribution $p(m_\alpha)$ at times $t=0$, $t_a/10$ and $t_a=3.75\times \pi\hbar^2/D^{(1)}$. (b) Corresponding plots  for an initial superposition of two wave packets of the form $\psi_\pm(\alpha)\propto\exp(-\cos(\alpha)^2/4\sigma_\alpha^2)$ with $\sigma_\alpha=0.06$. We choose $D^{(1)}=10\hbar^3/I$. Note how the fine angular momentum structures are quickly washed out already after the short time  $t_b=0.5\times\pi\hbar^2/D^{(1)}$.
	}
\end{figure}

Equation \eqref{eq:wignerpl} can be solved for arbitrary initial states $w_0(\alpha,m_\alpha)$. Since the special case $D^{(1)} = 0$ was discussed in the context of localization due to CSL \cite{schrinski2017}, we only discuss the case $D^{(2)} = 0$. The equation (\ref{eq:wignerpl}) can be solved by using the ansatz
\begin{equation}
\label{eq:planarwig}
w_t(\alpha,m_\alpha)=\sum_{\ell\in\ZZ}\int_{-\pi}^{\pi} \dint{\alpha'}w_0\left(\alpha-\frac{\hbar m_\alpha t}{I}-\alpha',m_\alpha-\ell \right)T_t(\alpha',\ell). 
\end{equation}
The kernel $T_t(\alpha',\ell)$ is obtained by solving the resulting difference equation with the method of characteristic functions,
\begin{equation}
T_t(\alpha',\ell)=\frac{\exp(-2D^{(1)}t/\hbar^2)}{2\pi}\sum_{k\in\ZZ}e^{ik(\alpha'+\ell\hbar t/2I)}I_\ell\left[\frac{2D^{(1)}t}{\hbar^2}\mathrm{sinc}\left(\frac{\hbar kt}{I}\right)\right],
\end{equation}
with $I_\ell(\cdot)$ the modified Bessel functions. It is easily observed that $T_t(\alpha',\ell)$ preserves the normalization of $w_t(\alpha,m_\alpha)$ and that it suppresses the rotational revivals at integer multiples of $\pi I/\hbar$ for $D^{(1)} > 0$.

In Fig. \ref{fig:planar}(a) we show the time evolution of the Wigner function for the ground state $w_0(\alpha,m_\alpha)=\delta_{m_\alpha 0}/2\pi$. The resulting momentum distribution can be calculated from Eq. \eqref{eq:planarwig} as
\begin{equation}\label{eq:pdistr}
p(m_\alpha) = \int_0^{2\pi}\rmd\alpha\, w_t(\alpha,m_\alpha)= \exp \left ( - \frac{2 D^{(1)} t}{\hbar^2} \right )I_{m_\alpha} \left ( \frac{2 D^{(1)} t}{\hbar^2} \right ),
\end{equation}
and it clearly shows angular momentum diffusion. The mean kinetic energy increases linearly with time, $\left \langle \sH_{\rm free} \right \rangle = D^{(1)}t/I$.  In order to see that 
the distribution (\ref{eq:pdistr}) turns into a Gaussian in the semiclassical limit  one multiplies it by $1/\hbar$ and draws the limit $\hbar \to 0$ while keeping $p_\alpha = \hbar m_\alpha$ constant.

Figure \ref{fig:planar}(b) displays the time evolution of a superposition of two orientational wave packets. One observes that the orientational coherences represented by the oscillatory structure around $\alpha=0$ rapidly vanish under the dynamics of the master equation,
while the free shearing motion of the individual wave packets is much less affected.

\section{Microscopic Derivation of Diffusion Constants}\label{sec:micro}

In this section we provide a microscopic derivation of the angular momentum diffusion tensors for two experimentally relevant scenarios, atom-molecule scattering in the Born approximation and Rayleigh-Gans scattering of photons.

\subsection{Atom-molecule scattering in the Born approximation}

A massive molecule rotating in a homogeneous mono-atomic gas at temperature $T$ interacts with the individual atoms via the general orientation-dependent interaction potential \cite{Stonebook}
\begin{equation}\label{eq:intpot}
V({\bf r},\Omega) = \sum_{\ell = 0}^\infty \sum_{m = -\ell}^\ell V_{\ell m}(r) Y_{\ell m}\left [\sR^{\rm T}(\Omega){\bf e}_r \right ],
\end{equation}
where ${\bf r} = r {\bf e}_r$ denotes the atom-molecule separation. If the interaction is weak in comparison to the energy of the impinging atom, the scattering amplitude can be evaluated in the Born approximation \cite{taylor},
\begin{equation}\label{eq:ampborn}
f(p{\bf n}', p{\bf n}; \Omega) \simeq \sum_{\ell =0}^\infty \sum_{m = -\ell}^\ell f_{\ell m}(p,\theta) Y_{\ell m}\left [ \sR^{\rm T} (\Omega) \frac{{\bf n} - {\bf n}'}{\vert {\bf n} - {\bf n'} \vert} \right ],
\end{equation}
where the expansion coefficients are determined by
\begin{equation}\label{eq:scattampcoeff}
f_{\ell m}(p,\theta) = -\frac{2 m i^\ell}{\hbar^2} \int_0^\infty \dint{r} r^2 V_{\ell m}(r) j_\ell \left [ \frac{2 p r}{\hbar} \sin \left ( \frac{\theta}{2} \right ) \right],
\end{equation}
with $j_\ell(\cdot)$ the spherical Bessel functions. The scattering amplitude \eqref{eq:ampborn} depends on the molecule's orientation only through its dependence on $({\bf n} - {\bf n}')/\vert{\bf n} - {\bf n}' \vert$. For $\ell = 0$ one obtains the well-known Born-approximation scattering amplitude for spherically symmetric potentials \cite{taylor}.

In the case of a weakly anisotropic interaction potential only low orders of $\ell$ contribute to Eqs.\ \eqref{eq:intpot} and \eqref{eq:ampborn} and we can approximate $f_{\ell m}(p,\theta) \simeq 0$ for $\ell > 2$. The diffusion coefficients \eqref{eq:diffconst} can thus be expressed in terms of the expansion coefficients $f_{\ell m}(p,\theta)$ by collecting them in the vector ${\bf A}_0(p,\theta)$ and the symmetric tensor $\sB_0(p,\theta)$,
\begin{equation}
{\bf A}_0 = \sqrt{\frac{3}{8 \pi}}  \left ( \begin{array}{c}
f_{1-1} - f_{11}  \\
-i ( f_{11} + f_{1-1}) \\
\sqrt{2} f_{10}
\end{array} \right )
\end{equation}
and
\begin{equation}
\fl
\sB_0 = \sqrt{\frac{15}{32 \pi}} \left ( \begin{array}{ccc}
f_{22} + f_{2-2} - \sqrt{\frac{2}{3}} f_{20} & i (f_{22}- f_{2-2}) & f_{2-1} - f_{21} \\
i (f_{22}- f_{2-2}) & -f_{2-2} - f_{22} - \sqrt{\frac{2}{3}} f_{20} & -i (f_{21} + f_{2-1}) \\
f_{2-1} - f_{21} & -i (f_{21} + f_{2-1}) & \sqrt{\frac{8}{3}} f_{20}
\end{array}
\right ),
\end{equation}
so that ${\bf A}(\Omega) = \sR(\Omega){\bf A}_0$ and $\sB(\Omega) = \sR(\Omega) \sB_0 \sR^{\rm T}(\Omega)$. Here we dropped the dependence on $(p,\theta)$ for better readability. Since the potential \eqref{eq:intpot} is real, the expansion vector $\sA_0$ is imaginary while the tensor $\sB_0$ is real.

As a simple example consider the azimuthally symmetric interaction potential
\begin{equation}\label{eq:pot2}
V({\bf r},\Omega) \simeq v(r) \left [ 1 + a_1 {\bf m}(\Omega) \cdot {\bf e}_r + \frac{\sqrt{5}a_2}{2} [{\bf m}(\Omega) \cdot {\bf e}_r ]^2 \right ],
\end{equation}
where the dimensionless constants $a_{1,2}$ quantify the anisotropy. Note that the dipole-induced-dipole interaction is of this form \cite{Stonebook}. Evaluating the coefficients \eqref{eq:scattampcoeff} with (\ref{eq:pot2}) one obtains a diagonal matrix $\sB_0$. Due to the azimuthal symmetry, the master equation is of the form (\ref{eq:supoplrlin}), (\ref{eq:supoplrquad}), characterized by the diffusion constants (\ref{eq:diffconstsym}). For thermally distributed gas momenta at temperature $T$ one arrives at the diffusion constants
\begin{eqnarray}
D^{(1,2)} & = & \sqrt{2 \pi m k_{\rm B} T}\frac{32 n_{\rm g} m a_{1,2}^2}{3 \hbar^2} \int_0^\infty \mathrm{d} \xi \xi e^{-\xi^2} g_{1,2}^2\left ( \sqrt{2 m k_{\rm B} T} \frac{2 \xi}{\hbar} \right ),
\end{eqnarray}
with the spherical form factor of the interaction potential
\begin{equation}
g_{\ell}(k) = \int_0^\infty \dint{r} r^2 v(r) j_{\ell}(k r).
\end{equation}
Hence, the diffusion constants increase quadratically with the anisotropy of the interaction potential.

\subsection{Light scattering}

In quantum experiments with optically levitated nanoparticles, scattering of stray photons is one of the most important sources of environment-induced momentum diffusion \cite{chang2010,romeroisart2010,romeroisart2011b}. Here, we consider a particle of volume $V_0$ incoherently illuminated by unpolarized, monochromatic radiation of wavenumber $k$. For simplicity, we assume that the particle extension is small compared to the photon wavelength so that the internal polarization field is completely determined by the susceptibility tensor $\chi_0$ with eigenvalues $\chi_i$.

The scattering amplitude can be calculated within the Rayleigh-Gans approximation \cite{jackson1999}. For linear rotors, the resulting decoherence rate has already been derived in Ref. \cite{stickler2016b}. Generalizing this treatment to arbitrary susceptibility tensors shows that the orientational localization rate of the nanoparticle is of the symmetric form \eqref{eq:difflocratequad}. The three diffusion coefficients are given by
\begin{eqnarray} \label{diffrg}
D^{(2)}_{i}&=\frac{\varepsilon_0 \hbar V_0^2 E_0^2 k^3}{36\pi}(\chi_j-\chi_k)^2,
\end{eqnarray}
with $(i,j,k)$ cyclic permutations of $(1,2,3)$ and $E_0$ the electric field amplitude per photon.

The angular momentum diffusion rates of an anisotropic nanoparticle thus depend  on its susceptibility anisotropies, which also determine the timescale on which orientational superpositions decohere \cite{stickler2016b,zhong2016}.
Note also that there is no $\cL_1$-contribution to photon-induced angular momentum diffusion, implying that Rayleigh-Gans scattering cannot completely localize the nanoparticle orientation. This is due to the fact that the photon wavelength is considered  much larger than the nanoparticle extension.

\section{Conclusion} \label{sec:conc}

The theory presented in this article provides a minimal model of complete orientational localization and quantum angular momentum diffusion of an arbitrarily shaped particle. We derived the associated master equation from the microscopic scattering processes responsible for orientational decoherence, studied its properties, and verified that its semiclassical limit coincides with the classical rotational diffusion equation. We  discussed the simplified equations resulting for symmetric particles, and specified the diffusion tensors for a number of relevant decoherence sources.

The current framework will be instrumental for the interpretation of future optomechanical experiments with non-spherical nanoparticles. Rotations give rise to features not familiar from center-of-mass motion, such as discrete angular momenta and orientational revivals. 
Observation and interpretation of such effects will require a quantitative understanding of the environmental influence. We illustrated this by discussing the decoherence-induced suppression of rotational revivals of the planar rotor.

The theory of quantum angular momentum diffusion may also be used to describe orientational localization effects in atom-molecule scattering experiments \cite{bennewitz1964} or matter wave interferometry with large molecules \cite{hornberger2012,eibenberger2013}. Our derivation directly relates the diffusion coefficients to the microscopic interaction potential between the impinging atom and the molecule. Finally, the quantum theory of angular momentum diffusion constitutes a first step toward a quantum description of the rotational friction and thermalization of a single object interacting with its environment.

\section*{References}

\providecommand{\newblock}{}

\end{document}